\documentclass[twocolumn,showpacs,preprintnumbers,amsmath,amssymb,times,pre]{revtex4}

\usepackage{graphicx}
\usepackage{dcolumn}
\usepackage{bm}
\usepackage{amsmath}
\usepackage{color}

\begin{document}

\title{Suppressing traffic-driven epidemic
spreading by adaptive routing strategy}

\author{Han-Xin Yang$^{1}$}\email{yanghanxin001@163.com}
\author{Zhen Wang$^{2}$}\email{zhenwang0@gmail.com}

\affiliation{$^{1}$Department of Physics, Fuzhou University, Fuzhou
350116, PR China}

\affiliation{$^{2}$Interdisciplinary Graduate School of Engineering
Sciences, Kyushu University, Kasuga-koen, Kasuga-shi, Fukuoka
816-8580, Japan}

\begin{abstract}
The design of routing strategies for traffic-driven epidemic
spreading has received increasing attention in recent years. In this
paper, we propose an adaptive routing strategy that incorporates
topological distance with local epidemic information through a
tunable parameter $h$. In the case where the traffic is free of
congestion, there exists an optimal value of routing parameter $h$,
leading to the maximal epidemic threshold. This means that epidemic
spreading can be more effectively controlled by adaptive routing,
compared to that of the static shortest path routing scheme.
Besides, we find that the optimal value of $h$ can greatly relieve
the traffic congestion in the case of finite node-delivering
capacity. We expect our work to provide new insights into the
effects of dynamic routings on traffic-driven epidemic spreading.
\end{abstract}

\date{\today}


\maketitle

\section{Introduction}

Epidemic spreading~\cite{1,2,3,4,5,6,7,8,9,10,11,12,13,14} and
traffic dynamics~\cite{16,17,18,19,20,21,22} on complex
networks~\cite{23,24,25} have attracted much attention in the past
decade. For a long time, the two types of dynamical processes have
been studied independently. However, epidemic spreading often
depends on traffic transportation. For example, a computer virus can
spread over Internet via data transmission. Another example is that
air transport tremendously accelerates the propagation of infectious
diseases among different countries.

The first attempt to incorporate traffic into epidemic spreading is
based on metapopulation model~\cite{m1}. This framework describes a
set of spatially structured interacting subpopulations as a network,
whose links denote the traveling path of individuals across
different subpopulations. Each subpopulation consists of a large
number of individuals. An infected individual can infect other
individuals in the same subpopulation. The metapopulation model is
often used to simulate the spread of human and animal diseases (such
as SARS and H1N1) among different cities. In a recent work, Meloni
$et$ $al.$ proposed another traffic-driven epidemic spreading model
which can be applied to study the propagation of computer virus on
the Internet~\cite{Meloni}. In Meloni model, each node of a network
represents a router on the Internet and the epidemic can spread
between nodes by the transmission of packets. A susceptible node
will be infected with some probability every time it receives a
packet from an infected neighboring node.

Meloni model has received increasing attention in recent
years~\cite{Meloni1,Meloni2,Meloni3,Meloni4,Meloni5,Meloni6}. It has
been found that the routing strategy can greatly effects epidemic
spreading~\cite{routing1,routing2}. Three routing algorithms have
been used in Meloni model. The first is the shortest-path routing
algorithm. The second is the local routing protocol~\cite{wang}, in
which each node does not know the whole network's topological
information and the packet is forwarded to a neighboring node $i$
with a probability that is proportional to the power of $i$'s
degree. The third is the efficient routing protocol~\cite{yan}, in
which each node in a network is assigned a weight that is
proportional to the power of its degree and The efficient path
between any two nodes corresponds to the route that makes the sum of
the nodes' weight (along the path) minimal.

All the above routing strategies are based on the network structure
and packets follow the fixed routes for a given network. In this
paper, we propose an adaptive routing strategy that integrates
topological distance with local epidemic information through a
tunable parameter $h$. In the adaptive routing strategy, a packet
can timely adjust its route according to the epidemic information of
its neighbors. Interestingly, we find that there exists an optimal
value of $h$, leading to the maximal epidemic threshold.

The paper is organized as follows. In Sec.~\ref{sec:model}, we
formalize the problem by introducing the adaptive routing strategy
into traffic-driven epidemic spreading. In Sec.~\ref{sec:infinite}
and Sec.~\ref{sec:finite}, we present the results for infinite and
finite node-delivering capacity respectively. Finally, we give a
conclusion in Sec.~\ref{sec: conclusion}.

\section{Model}\label{sec:model}

Following the work of Meloni $et$ $al.$~\cite{Meloni}, we
incorporate the traffic dynamics into the classical
susceptible-infected-susceptible model~\cite{SIS} of epidemic
spreading as follows.

(i) {\em Adaptive routing protocol.} In a network of size $N$, at
each time step, $\lambda N$ new packets are generated with randomly
chosen sources and destinations (we call $\lambda$ as the
packet-generation rate), and each node can deliver at most $C$
packets towards their destinations. To deliver a packet to its
destination, a node performs a local search within its neighbors. If
the packet's destination is found inside the searched area, it will
be delivered directly to the destination. Otherwise, the packet is
forwarded to a neighboring node $i$ toward its destination $j$ with
the smallest value of effective distance, denoted by
\begin{equation}
d_{eff}^{ij}=h\cdot D_{ij}+(1-h)\delta_{i},
 \label{5}
\end{equation}
where $h$ is the routing parameter ($0\leq h\leq1$), $D_{ij}$ is the
topological distance between nodes $i$ and $j$, and $\delta_{i}=1$
($\delta_{i}=0$) if node $i$ is infected (uninfected) in the
previous time step.

It is worth noting that when $h = 1$, the adaptive routing recovers
to the traditional shortest path routing. Once a packet reaches its
destination, it is removed from the system. The queue length of each
node is assumed to be unlimited and the first-in-first-out principle
holds for the queue.

(ii) {\em Epidemic dynamics.} After a transient time, the total
number of delivered packets at each time will reach a steady value,
then an initial fraction of nodes $\rho_{0}$ is set to be infected
(e.g., we set $\rho_{0}=0.1$ in numerical experiments). The
infection spreads in the network {\em through packet exchanges}.
Each susceptible node has the probability $\beta$ of being infected
every time it receives a packet from an infected neighbor. The
infected nodes recover at rate $\mu$ (we set $\mu=1$ in this paper).

In the following, we carry out simulations systematically by
employing traffic-driven epidemic spreading on the
Barab\'{a}si-Albert (BA) scale-free networks~\cite{ba}. The size of
the BA network is set to be $N=2000$ and the average degree of the
network is $\langle k \rangle=4$. Each data point results from an
average over 30 different realizations.

\section{Results for infinite node-delivering capacity}\label{sec:infinite}

In the case where the node-delivering capacity is infinite
($C\rightarrow \infty$), traffic congestion will not occur in the
network.

\begin{figure}
\begin{center}
\scalebox{0.2}[0.2]{\includegraphics{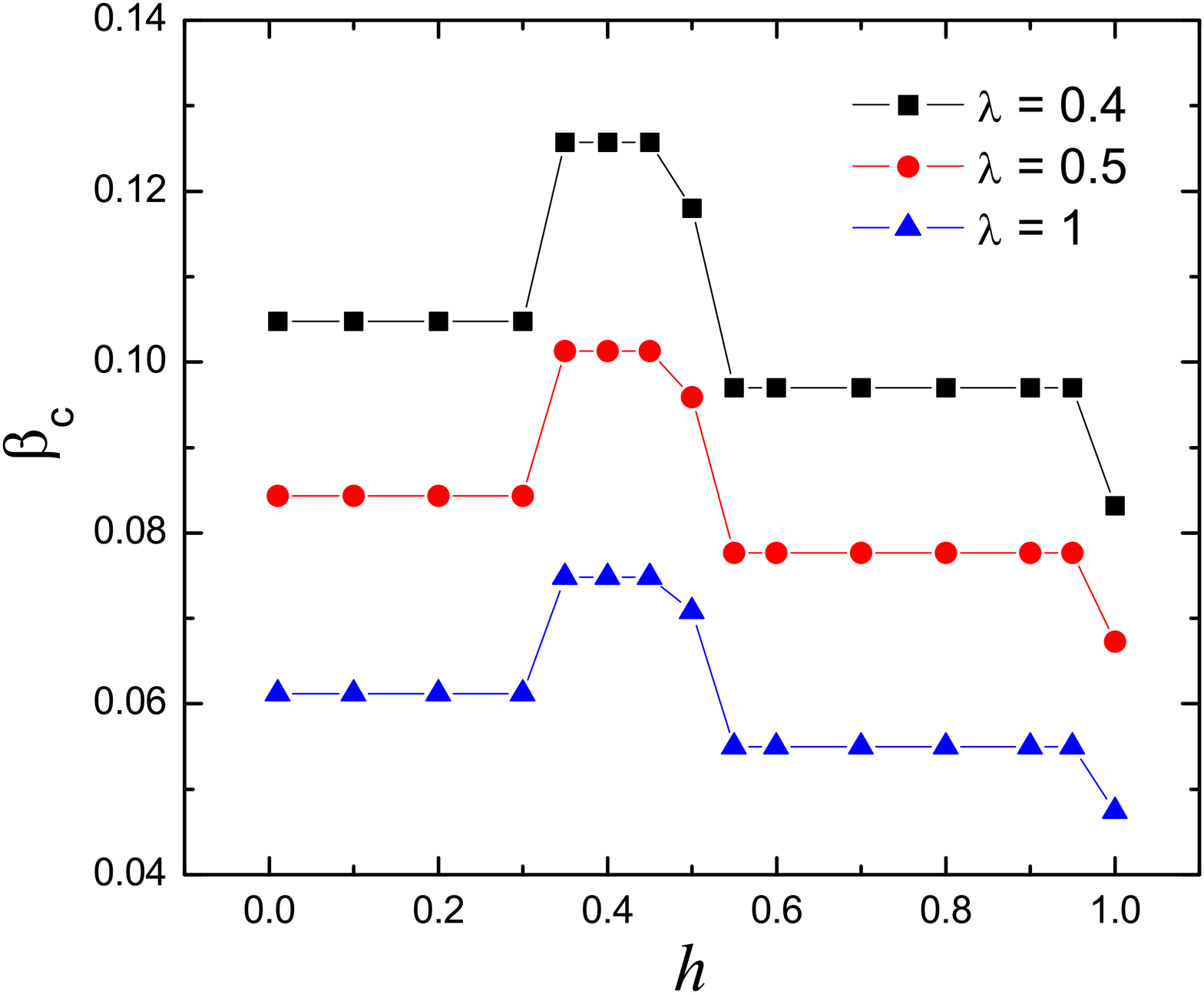}} \caption{(Color
online) The epidemic threshold $\beta_{c}$ as a function of the
routing parameter $h$ for different values of the packet-generation
rate $\lambda$. The node-delivering capacity is infinite.}
\label{fig1}
\end{center}
\end{figure}

Previous studies have shown that there exists an epidemic threshold
$\beta_{c}$, below which the epidemic goes extinct~\cite{Meloni}.
Figure~\ref{fig1} shows the dependence of $\beta_{c}$ on $h$ for
different values of the packet-generation rate $\lambda$. We find
that for each value of $\lambda$, there exists an optimal region of
$h$ (around 0.4), leading to the maximum $\beta_{c}$. This
phenomenon indicates that the integration of topological structure
and epidemic information can effectively suppress the outbreak of
epidemic.

\begin{figure}
\begin{center}
\scalebox{0.2}[0.2]{\includegraphics{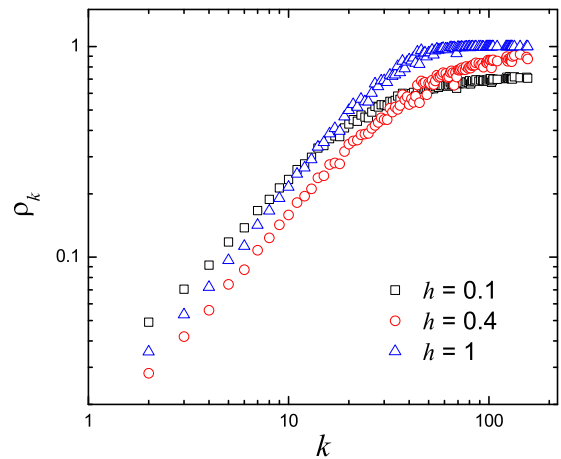}} \caption{(Color
online) The density of infected nodes $\rho_{k}$ as a function of
the degree $k$ for different values of the routing parameter $h$.
The packet-generation rate $\lambda=0.5$ and the spreading rate
$\beta=0.13$. The node-delivering capacity is infinite.}
\label{fig3}
\end{center}
\end{figure}

To understand the emergence of the optimal $h$, we study the density
of infected nodes $\rho_{k}$ as a function of the degree $k$ for
different values of the routing parameter $h$ when the
packet-generation rate $\lambda=0.5$ and the spreading rate
$\beta=0.13$. From Fig.~\ref{fig3}, one can see that $\rho_{k}$
increases as $k$ increases for each value of $h$, indicating that
larger-degree nodes are more likely to be infected. As $h$ decreases
from 1 to 0.1, the infection probability for large-degree nodes
(e.g., $k>50$) decreases while small-degree nodes (e.g., $k<10$) are
more likely to be infected.

For $h=1$, many shortest paths go through large-degree nodes,
leading to a heavy traffic load and a high infection probability for
large-degree nodes. For $h<1$, packets can bypass these infected
large-degree nodes, which reduces the infection probability for
large-degree nodes. However, for too small value of $h$,
small-degree become more likely to be infected since packets reroute
via small-degree nodes more frequently. For the moderate value of
$h$, epidemic can simultaneously die out in both large and small
degree classes, leading to the maximal epidemic threshold.

\begin{figure}
\begin{center}
\scalebox{0.2}[0.2]{\includegraphics{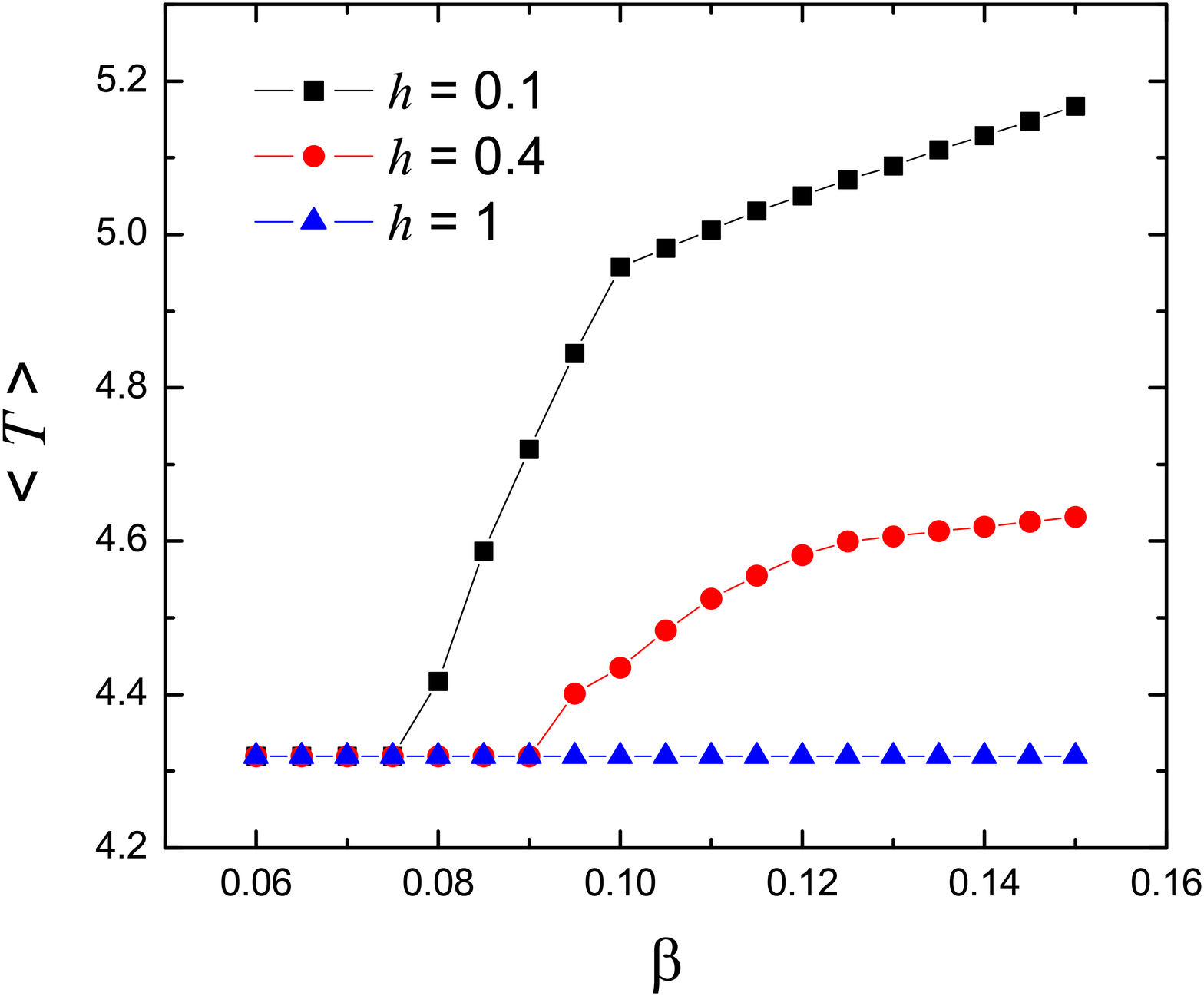}} \caption{(Color
online) The average traveling time of a packet $\langle T \rangle$
as a function of the spreading rate $\beta$ for different values of
the routing parameter $h$. The packet-generation rate $\lambda=0.5$.
The node-delivering capacity is infinite.}\label{fig4}
\end{center}
\end{figure}

\begin{figure}
\begin{center}
\scalebox{0.2}[0.2]{\includegraphics{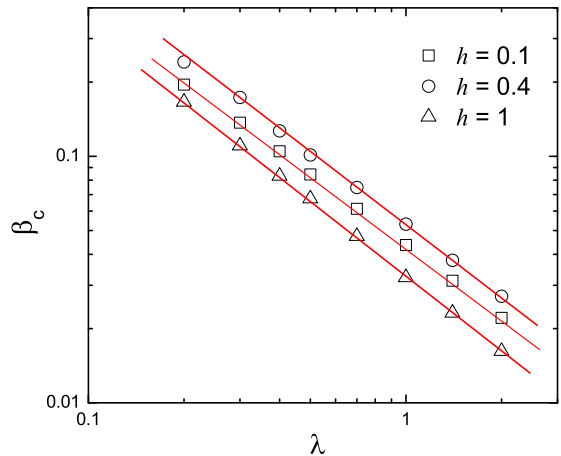}} \caption{(Color
online) The epidemic threshold $\beta_{c}$ as a function of the
packet-generation rate $\lambda$ for different values of the routing
parameter $h$. The slopes of the fitted lines are about -1. The
node-delivering capacity is infinite. }\label{fig5}
\end{center}
\end{figure}

Figure~\ref{fig4} shows the average traveling time of a packet
$\langle T \rangle$ as a function of the spreading rate $\beta$ for
different values of the routing parameter $h$. One can see that for
the shortest path routing ($h=1$), $\langle T \rangle$ is
independence of $\beta$. For $h<1$, $\langle T \rangle$ keeps
unchanged when $\beta<\beta_{c}$ while $\langle T \rangle$ increases
with $\beta$ when $\beta>\beta_{c}$. This is because when
$\beta<\beta_{c}$, epidemic dies out in the network and all packets
are delivered along the shortest path. When $h<1$ and
$\beta>\beta_{c}$, packets bypass infected large-degree nodes and
reroute via small-degree nodes, leading to a longer traveling time.
From Fig.~\ref{fig4}, we can also observe that for a fixed value of
$\beta$ (e.g., $\beta=0.12$), $\langle T \rangle$ increases as $h$
decreases from 1 to 0.1.

Figure~\ref{fig5} shows the epidemic threshold $\beta_{c}$ as a
function of the packet-generation rate $\lambda$ for different
values of the routing parameter $h$. One can see that for each value
of $h$, $\beta_{c}$ scales inversely with $\lambda$, indicating that
the increase of traffic flow facilitates the outbreak of epidemic.
The similar result has also been found in Ref.~\cite{Meloni}.

\section{Results for finite node-delivering capacity}\label{sec:finite}

When the node-delivering capacity is finite, traffic congestion can
occur if the packet-generating rate exceeds a critical value
$\lambda_{c}$~\cite{alg1,alg2}. The traffic throughput of a network
can be characterized by the critical value $\lambda_{c}$.

\begin{figure}
\begin{center}
\scalebox{0.2}[0.2]{\includegraphics{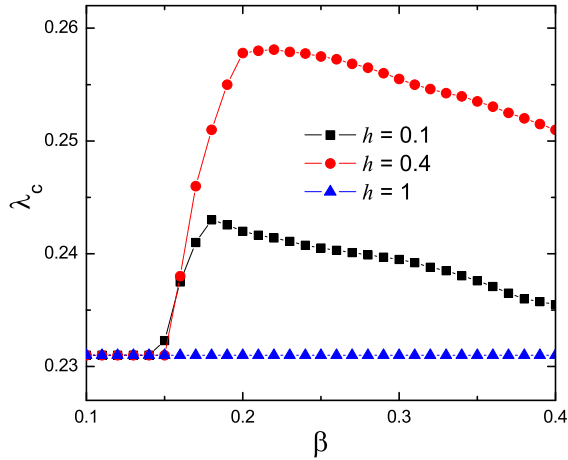}} \caption{(Color
online) The critical packet-generating rate $\lambda_{c}$ as a
function of the spreading rate $\beta$ for different values of the
routing parameter $h$. The node-delivering capacity is $C=100$. }
\label{fig6}
\end{center}
\end{figure}

Figure~\ref{fig6} shows the critical packet-generating rate
$\lambda_{c}$ as a function of the spreading rate $\beta$ for
different values of the routing parameter $h$. One can see that for
$h=1$, $\lambda_{c}$ keeps unchanged for different values of
$\beta$. For $h=0.1$ or $h=0.4$, there exists an optimal value of
$\beta$, leading to the maximum $\lambda_{c}$. This non-monotonic
relationship can be explained as follows. For $\beta<\beta_{c}$,
epidemic dies out and all packets are delivered along the shortest
path, leading to a heavy load on large-degree nodes. When $\beta$ is
a little larger than $\beta_{c}$, packets bypass infected
large-degree nodes, which reduces the traffic load of large-degree
nodes and enhances the traffic throughput of the network. However,
for too large value of $\beta$, epidemic spreads so widely in the
network that packets reroute many times, which increases the average
traveling time of a packet and reduces the traffic throughput of the
network.

\begin{figure}
\begin{center}
\scalebox{0.2}[0.2]{\includegraphics{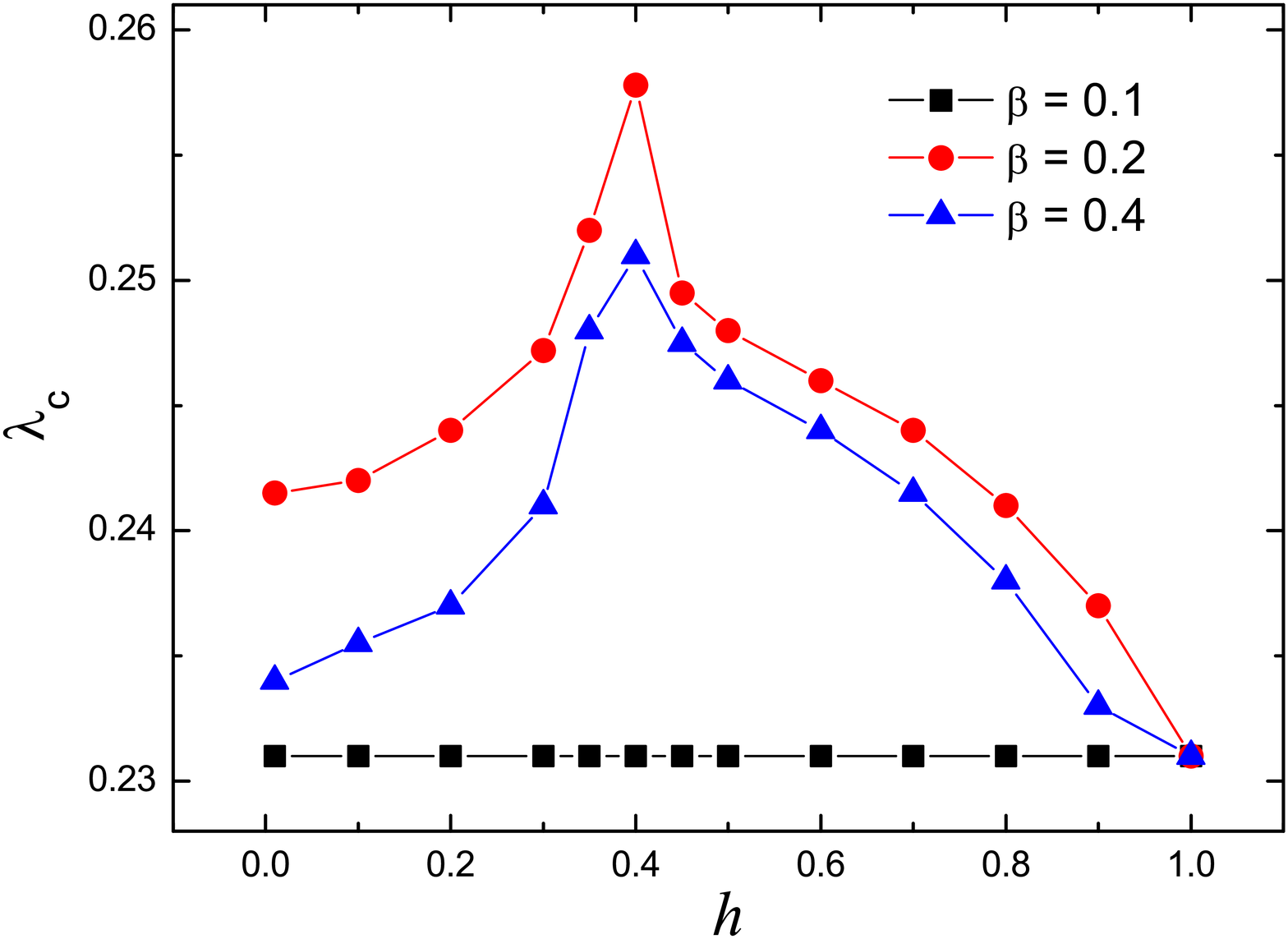}} \caption{(Color
online) The critical packet-generating rate $\lambda_{c}$ as a
function of the routing parameter $h$ for different values of the
spreading rate $\beta$. The node-delivering capacity is $C=100$.}
\label{fig7}
\end{center}
\end{figure}

\begin{figure}
\begin{center}
\scalebox{0.2}[0.2]{\includegraphics{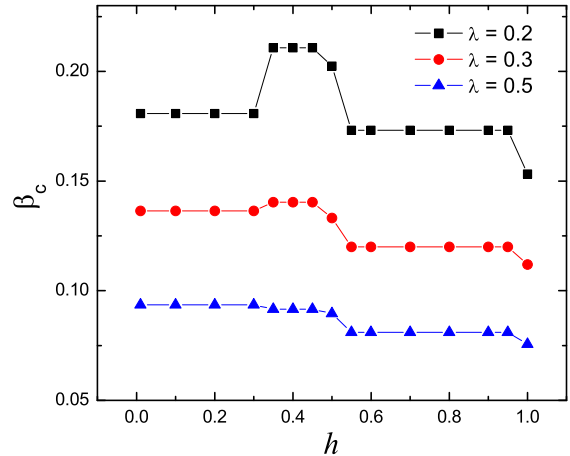}} \caption{(Color
online) The epidemic threshold $\beta_{c}$ as a function of the
routing parameter $h$ for different values of the packet-generation
rate $\lambda$. The node-delivering capacity $C=100$. } \label{fig8}
\end{center}
\end{figure}

Figure~\ref{fig7} shows the dependence of $\lambda_{c}$ on $h$ for
different values of $\beta$. One can see that for the small value of
$\beta$ (e.g., $\beta=0.1$), $\lambda_{c}$ is independence of $h$
since epidemic dies out and all packets are delivered along the
shortest path. For $\beta=0.2$ or $\beta=0.4$, $\lambda_{c}$
maximized at $h=0.4$. Note that the optimal value of $h$ resulting
in the maximum $\lambda_{c}$ for finite node-delivering capacity and
the maximum $\beta_{c}$ for infinite node-delivering capacity is
almost the same.

Figure~\ref{fig8} shows the epidemic threshold $\beta_{c}$ as a
function of the routing parameter $h$ for different values of the
packet-generation rate $\lambda$ when the node-delivering capacity
$C=100$. One can observe that for small values of $\lambda$ (e.g.,
$\lambda=0.2$), there also exists an optimal region of $h$ (around
0.4), leading to the maximal $\beta_{c}$. However, when $\lambda$ is
large (e.g., $\lambda=0.5$), $\beta_{c}$ decreases with the increase
of $h$.

\section{Conclusions}\label{sec: conclusion}

In conclusion, we have proposed an adaptive routing strategy which
incorporates topological distance with local epidemic information
through a tunable parameter $h$. For $h=1$, the adaptive routing is
reduced to the shortest path routing. Compared to static routings,
in adaptive routing packets can change their routing paths when
epidemic outbreaks.

Our main findings are as follows. (i) In the case of infinite
node-delivering capacity, the epidemic threshold is maximized at
about $h=0.4$. (ii) In the case of finite node-delivering capacity,
the epidemic threshold is maximized at about $h=0.4$ when the
packet-generation rate is small but the epidemic threshold deceases
as $h$ increases when the packet-generation rate is large. (iii) In
the case of finite node-delivering capacity, the traffic throughput
of the network is maximized at about $h=0.4$.

It is interesting to note that in static routings such as the
shortest path routing, epidemic spreading has no effect on the
traffic throughput of the network. However, in adaptive routing,
epidemic spreading and traffic transportation interact with each
other. Through the rerouting of packets, both the infection
probability and the traffic load of large-degree nodes can be
reduced. Our results can provide insights into devising effective
routing strategies to suppress the spreading of computer virus on
the Internet.

\begin{acknowledgments}
This work was supported by the National Natural Science Foundation
of China under Grant No. 6140308 and the training plan for
Distinguished Young Scholars of Fujian Province, China.
\end{acknowledgments}

\end{document}